\documentclass[12pt,english]{article}
\usepackage[T1]{fontenc}
\usepackage[latin1]{inputenc}
\usepackage{a4wide}
\usepackage{amsmath}
\usepackage{color}
\usepackage{graphicx}
\usepackage{amssymb}

\makeatletter

\providecommand{\tabularnewline}{\\}

\usepackage{amssymb}

\usepackage{babel}
\makeatother
\begin{document}

\title{Exact finite size spectrum in super sine-Gordon model}

\author{\'{A}. Heged\H{u}s$^{1,2}$, F. Ravanini$^{3,1}$ and J. Suzuki$^{4}$}

\maketitle
\begin{center}\emph{$^{1}$I.N.F.N., Sezione di Bologna - Italy}\end{center}

\begin{center}\emph{$^{2}$Research Institute for Particle and Nuclear
Physics,}\\
\emph{Hungarian Academy of Sciences, Budapest - Hungary}\end{center}

\begin{center}\emph{$^{3}$Dip. di Fisica, Università di Bologna -
Italy}\end{center}

\begin{center}\emph{$^{4}$Dept. of Physics, Shizuoka University -
Japan}\end{center}

\begin{abstract}
We present the coupled nonlinear integral equations (NLIE) governing
the finite size effects in N=1 super sine-Gordon model for the vacuum
as well as for the excited states. Their infrared limit correctly
yields the scattering data of the super sine-Gordon S-matrix conjectured
by Ahn \cite{ahn}. Ultraviolet analysis is in agreement with the
expected conformal data of $c=3/2$ CFT. Conformal perturbation theory
further corroborates this result.
\end{abstract}

\section{Introduction}

Finite Size Effects in two dimensional integrable QFT can be studied
exactly by means of sets of coupled nonlinear integral equations that
can be thought as continuum versions of Bethe Ansatz equations. In
particular, the equation introduced for the Sine-Gordon (SG) model
by Destri and De Vega \cite{DdV1,DdV2} through a light-cone lattice
regularization and basically equivalent to a similar one found by
Kl\"{u}mper, Batchelor and Pearce \cite{KBP,KP} for the spin 1/2
XXZ chain, once extended to excited states \cite{FMQR,DdV3}, has
proven to be a very important tool in establishing and testing the
correct connection between the Factorized Scattering description and
the Perturbed CFT formulation of SG model \cite{FRT1,FRT2,FRT3}.
Restricted SG models can be put in correspondence with minimal CFT
perturbed by their $\phi_{13}$ operators in a similar way \cite{FRT4}.

It would be a result of principal importance to be able to formulate
such non-linear integral equations (NLIE) for many other integrable
QFTs. In this paper we report the construction of NLIEs for the N=1
Super Sine-Gordon (SSG) model. The importance of this generalization
lies in the manifest supersymmetry of this integrable model and the
link it has with applications in various areas of physics ranging
from condensed matter to strings.

To get the NLIEs for SSG model, we proceed by analogy with the pure
SG case. There, the NLIE was obtained through a light-cone lattice
construction of a 6-vertex model, which is known to be equivalent,
from the Bethe Ansatz point of view, to an alternating inhomogeneous
spin 1/2 XXZ chain. In the continuum limit this lattice model is shown
to give the equations of motion of SG (or equivalently massive Thirring)
model \cite{DdV0,DS}. It is known since a long time that the 6-vertex
model on a square lattice, or equivalently spin 1/2 \emph{homogeneous}
XXZ chain, renormalizes on the continuum to a free massless boson,
i.e. $c=1$ CFT \cite{boson}. The analog model with spin 1, the integrable
spin 1 homogeneous integrable XXZ chain is equivalent to the square
lattice 19-vertex model and renormalizes to a $c=3/2$ boson + Majorana
fermion theory which is explicitly N=1 supersymmetric.

It is therefore natural to conjecture that the full SSG model can
be constructed starting from a light-cone lattice 19-vertex model,
or equivalently from a spin 1 alternating inhomogeneous integrable
XXZ chain \cite{resh-sal}. A direct deduction of the equations of
motion, like in \cite{DdV0} or \cite{DS} is too cumbersome and beyond
the scope of the present paper.

The Bethe Ansatz for spin 1 integrable XXZ chain has been written
long time ago \cite{bethe-ansatz} and can be generalized with the
introduction of an alternating inhomogeneity trivially. In the homogeneous
case, a coupled set of two NLIE's equivalent to it on the lattice
has been deduced, by use of transfer matrix recurrence relations and
T-Q formalism in ref.\cite{suzuki}, including excited states. Their
generalization to the inhomogeneous case is also trivial. On the other
hand a system of two NLIE was conjectured on the continuum by C. Dunning
\cite{dunning} for the vacuum state and checked to give the expected
ultraviolet central charge $c=3/2$. Further analysis of Finite Size
Effects in SSG model was carried out in \cite{Gabor1}.

In section \ref{sec:SSGmodel} we recall the setup of the model in
Lagrangian formulation, as a perturbed CFT and a factorized scattering
theory. In section \ref{sec:XXZ} we present the alternating inhomogeneous
integrable spin 1 XXZ chain and its Bethe ansatz. The NLIE is presented
in section \ref{sec:NLIE} for the spin chain on the lattice and,
by performing the scaling limit, in the continuum. Section \ref{sec:IR}
is devoted to the infrared (IR) analysis reconstructing the supersymmetric
solitonic S-matrix. From this section on we limit ourselves mostly
to the repulsive regime. The interesting features of the attractive
regime, as well as the implications for superminimal model perturbed
by $\Phi_{13}$ are left for a future investigation. In section \ref{sec:UV}
the ultraviolet (UV) limit of the NLIE is computed and shown to reproduce
data compatible with the expected $c=3/2$ CFT. A numerical comparison
with predictions of conformal perturbation theory is presented in
section \ref{sec:PCFT}. Finally we draw our conclusions and perspectives
for future work in section \ref{sec:Conclusions}.

\section{The super sine-Gordon (SSG) model\label{sec:SSGmodel}}

The SSG model can be defined on a cylinder of circumference $L$ (thought
as spatial direction, with infinite time direction) by the action\[
\mathcal{A}_{SSG}=\int dt\int_{0}^{L}dx\int d\vartheta d\bar{\vartheta}\left(D\Phi\bar{D}\Phi+\mu\cos\beta\Phi\right)\]
 where $\beta$ is the coupling, $\mu$ a bare mass and the superfield
$\Phi$ expands in components as\[
\Phi(x,t,\theta,\bar{\theta})=\varphi(x,t)+\vartheta\psi(x,t)+\bar{\vartheta}\bar{\psi}(x,t)+\vartheta\bar{\vartheta}F(x,t)\]
 The action in components, once the auxiliary field $F$ is eliminated
by the Euler-Lagrange equations of motion, reads\begin{equation}
\mathcal{A}_{SSG}=\int dt\int_{0}^{L}dx\left(\frac{1}{2}\partial_{\mu}\varphi\partial^{\mu}\varphi+i\bar{\psi}\gamma^{\mu}\partial_{\mu}\psi+\mu\psi\bar{\psi}\cos\frac{\beta}{2}\varphi+\frac{\mu^{2}}{\beta^{2}}\cos\beta\varphi\right)\label{SSG-action}\end{equation}

This model is known to be integrable since a long time and evidently
possesses $N=1$ supersymmetry. Its ultraviolet (UV) limit is given
by a theory of a free massless boson $\Omega(x,t)=\phi(z)+\bar{\phi}(\bar{z})=\frac{\varphi}{\sqrt{4\pi}}$
(here $z=x+it$, $\bar{z}=x-it$) compactified on a circle of radius
$R=\frac{4\sqrt{\pi}}{\beta}$ (i.e. with quasi periodic boundary
conditions $\Omega(x+L,t)=\Omega(x,t)+2\pi mR$ at fixed time, $m$
being called the \emph{winding number}) and a free massless Majorana
fermion with antiperiodic (NS = Neveu-Schwarz) or periodic (R = Ramond)
boundary conditions $\psi(x+L,t)=\pm\psi(x,t)$. This CFT obviously
has central charge $c=\frac{3}{2}$ and a $(U(1)\times\mathbb{Z}_{2})_{L}\times(U(1)\times\mathbb{Z}_{2})_{R}$
symmetry and shows $N=1$ supersymmetry.

In the NS sector the primary fields are given by vertex operators\[
V_{n,m}^{(r,\bar{r})}(z,\bar{z})=\bar{\psi}_{\bar{r}}\psi_{r}:e^{i\left[\left(\frac{n}{R}+\frac{mR}{2}\right)\phi(z)+\left(\frac{n}{R}-\frac{mR}{2}\right)\bar{\phi}(\bar{z})\right]}:\]
 where $r,\bar{r}\in\{0,1\}$, $\psi_{0}=1$, $\psi_{1}=\psi$ of
conformal dimensions \[
\Delta_{n,m}^{(r,\bar{r})}=\frac{1}{2}\left(\frac{n}{R}+\frac{mR}{2}\right)^{2}+\frac{r}{2}\quad,\quad\bar{\Delta}_{n,m}^{(r,\bar{r})}=\frac{1}{2}\left(\frac{n}{R}-\frac{mR}{2}\right)^{2}+\frac{\bar{r}}{2}\]

In the R sector the primary fields are realized by bosonic vertex
operators multiplied by the Ising spin field\[
R_{n,m}(z,\bar{z})=\sigma(z,\bar{z}):e^{i\left[\left(\frac{n}{R}+\frac{mR}{2}\right)\phi(z)+\left(\frac{n}{R}-\frac{mR}{2}\right)\bar{\phi}(\bar{z})\right]}:\]
of conformal dimensions\[
\Delta_{n,m}=\frac{1}{2}\left(\frac{n}{R}+\frac{mR}{2}\right)^{2}+\frac{1}{16}\quad,\quad\bar{\Delta}_{n,m}=\frac{1}{2}\left(\frac{n}{R}-\frac{mR}{2}\right)^{2}+\frac{1}{16}\]

All the other secondary states in the theory can be obtained by applying
the Heisenberg algebra and the fermion algebra modes to the primaries.
See \cite{Gabor1} for details.

The main object encoding the UV operator content of the theory is
the modular invariant partition function of the $c=\frac{3}{2}$ CFT,
obtained in \cite{DSZ}. It reads\begin{eqnarray*}
Z(R) & = & \frac{1}{|\eta|^{2}}\left\{ (\chi_{0}\bar{\chi}_{1/2}+\chi_{1/2}\bar{\chi}_{0})\sum_{n\in\mathbb{Z}+\frac{1}{2},\, m\in2\mathbb{Z}+1}\right.\\
 & + & \left.(|\chi_{0}|^{2}+|\chi_{1/2}|^{2})\sum_{n\in\mathbb{Z},\, m\in2\mathbb{Z}}+|\chi_{1/16}|^{2}\sum_{2n-m\in2\mathbb{Z}+1}\right\} q^{\Delta_{n,m}^{(0,0)}}\bar{q}^{\bar{\Delta}_{n,m}^{(0,0)}}\end{eqnarray*}
where $\eta(q)$ is the Dedekind function and $q=e^{2\pi i\tau}$,
$\tau$ being the modular parameter. We see from the partition function
that in the NS sector the quantum numbers $n,m,r,\bar{r}$ describing
the operator content can take the values\[
n\in\mathbb{Z}+\frac{1}{2}\qquad m\in2\mathbb{Z}+1\qquad\left\{ \begin{array}{cc}
r=1\quad & \bar{r}=0\\
r=0\quad & \bar{r}=1\end{array}\right.\]
or\[
n\in\mathbb{Z}\qquad m\in2\mathbb{Z}\qquad\left\{ \begin{array}{c}
r=\bar{r}=0\\
r=\bar{r}=1\end{array}\right.\]
while in the R sector they are constrained to take the values\[
n\in\mathbb{Z}\qquad m\in2\mathbb{Z}+1\]
or\[
n\in\mathbb{Z}+\frac{1}{2}\qquad m\in2\mathbb{Z}\]

The SSG model can be seen as a perturbation of this $c=3/2$ CFT by
a relevant operator\begin{equation}
\Phi_{pert}=\sqrt{2}\pi\left(V_{1,0}^{(1,1)}+V_{-1,0}^{(1,1)}\right)=\sqrt{2}\pi:\psi\bar{\psi}\cos\frac{\beta}{2}\varphi:\label{eq:pert}\end{equation}
of conformal dimension $\Delta_{pert}=\bar{\Delta}_{pert}=\frac{1}{2}+\frac{\beta^{2}}{32\pi}$.
The other perturbing field\[
\frac{1}{2}\left(V_{(2,0)}^{(0,0)}+V_{(-2,0)}^{(0,0)}\right)=:\cos\beta\varphi:\]
necessary in the classical action to ensure supersymmetry, can be
shown to give a contribution to quantum correlators that goes to zero
with the cutoff \cite{Mansfield} and can be neglected in a quantum
formulation. See \cite{Gabor1} for a more detailed discussion of
these two perturbation terms.

The operator (\ref{eq:pert}) is relevant ($\Delta_{pert}<1$) if
the SSG coupling is in the range $0<\beta^{2}<16\pi$. At exactly
$\beta^{2}=16\pi$ it becomes marginally relevant and describes a
perturbation of the WZW $SU(2)_{k=2}$ model by its $J^{a}\bar{J}^{a}$
operator.

Opposite to this UV description of SSG as a perturbed CFT, there exists
a factorized scattering theory that has been conjectured by C. Ahn
\cite{ahn} to describe the asymptotic particle properties of SSG
model. The S-matrix is written in detail in the original paper \cite{ahn}
and the vacuum structure is carefully analyzed in \cite{Gabor1}.
While not repeating here those data, we just recall that the S-matrix
has a tensor product form\[
S(\theta)=S_{RSOS(2)}(\theta)\otimes S_{SG}(\theta,\beta_{SG})\]
where the RSOS part is the S-matrix of the tricritical Ising model
perturbed by $\phi_{13}$ and describes the supersymmetric structure,
while the SG part is the usual solitonic S-matrix of pure sine-Gordon
model, at $\beta_{SG}^{2}=\frac{16\pi\beta^{2}}{16\pi+\beta^{2}}$,
describing the bosonic degrees of freedom. The vacuum structure shows,
similarly to SG, infinitely many degenerate vacua periodically repeating.
The additional RSOS structure amounts to the fact that in each period
there are three degenerate vacua, labeled by $\{0,\frac{1}{2},1\}$
and connected by the adjacency of the $A_{3}$ Dynkin diagram. The
elementary excitations are supersymmetric solitons (kinks) that interpolate
these vacua \cite{Gabor1}. They are the only asymptotic particles
of the model if $\frac{16\pi}{3}<\beta^{2}\leq16\pi$ (\emph{repulsive}
regime). For $\beta^{2}<\frac{16\pi}{3}$ bound states (supersymmetric
breathers) appear (\emph{attractive} regime).

\section{The inhomogeneous spin 1 XXZ chain and its Bethe ansatz\label{sec:XXZ}}

As said in the introduction, we assume, by analogy with the SG case,
that the SSG model can be constructed by taking the appropriate continuum
limit of an integrable lattice model: the light-cone lattice 19-vertex
model. For a description of this lattice model including its Boltzmann
weights, see \cite{DSZ}. Actually, there are infinitely many possible
lattice models providing SSG as a continuum limit. Among them, it
is convenient for our purposes to choose one that preserves integrability
on the lattice, so that the Bethe ansatz machinery can be used to
diagonalize its transfer matrix and get the energy levels and in principle
other measurable quantities.

The square lattice 19-vertex model, in the limit of continuous time
direction while keeping the space direction discretized , is related
to the integrable homogeneous spin 1 XXZ chain \cite{FZ-spin1} with
anisotropy $\gamma$. Its continuum limit is a system of a free massless
boson compactified on a radius $R=\sqrt{\pi/(\pi-2\gamma)}$. The
light-cone lattice version of the same model is instead equivalent
to a spin 1 XXZ chain, with $N$ sites labeled by $n=1,...,N$, lattice
spacing $a$, anisotropy $\gamma$ and inhomogeneity $\lambda_{n}=(-1)^{n+1}\Theta$.
The relation between the compactification radius $R$ and the SSG
coupling $\beta$ implies that the anisotropy $\gamma$ is related
to $\beta$ through $\frac{\gamma}{\pi}=\Delta_{pert}(\beta)$. It
is often conveniently parametrized by $\gamma=\frac{\pi}{p+2}$. \textcolor{black}{Throughout
this paper, we will mainly focus on the range $\gamma<\frac{\pi}{3}$,
i.e. $p>1$. This corresponds to the repulsive regime of SSG model. }

We start from the description of the homogeneous system. At each site
$n$ of the chain a 3-dimensional space $V_{n}\sim\mathbb{C}^{3}$
is defined, on which the spin operators of the Hamiltonian act. Let
$R_{n,m}(\theta)$ be the $R$ matrix acting non trivially on $V_{n}\otimes V_{m}$.
Its explicit form can be found in \cite{FZ-spin1}.

The homogeneous transfer matrix $T_{2}(\theta)$ is defined by the
trace of the product of $R$ over the auxiliary space $V_{a}\sim\mathbb{C}^{3}$
\[
T_{2}(\theta):={\textrm{Tr}}_{V_{a}}R_{a,1}(\theta)\cdots R_{a,N}(\theta).\]
 Periodic boundary conditions are imposed here.

The hamiltonian of the associated spin 1 XXZ chain is identified with
${\mathcal{H}}=\frac{\pi}{\gamma}\frac{d}{d\theta}\log T_{2}(\theta)$.\begin{equation}
{\mathcal{H}}\negthickspace=\negthickspace\sum_{n=1}^{N}\left(\sigma_{n}^{\perp}-(\sigma_{n}^{\perp})^{2}+\cos2\gamma(\sigma_{n}^{z}-(\sigma_{n}^{z})^{2})-(2\cos\gamma-1)(\sigma_{n}^{\perp}\sigma_{n}^{z}+\sigma_{n}^{z}\sigma_{n}^{\perp})-4\sin^{2}\gamma(S_{n}^{z})^{2}\right)\label{hamil}\end{equation}
 where we write $\sigma_{n}=S_{n}\cdot S_{n+1}=\sigma_{n}^{\perp}+\sigma_{n}^{z}.$

Next we consider the inhomogeneous generalization \cite{DdV1,DdV2,resh-sal},
($N\in2\mathbb{Z}$)\[
T_{2}(\theta|\{\lambda_{j}\}):={\textrm{Tr}}_{V_{a}}R_{a,1}(\theta-\lambda_{1})\cdots R_{a,N}(\theta-\lambda_{N}).\]
 By following the light-cone approach, we adopt a particular \emph{alternating}
choice of the inhomogeneities $\lambda_{n}=(-1)^{n}\Theta$ and write
$T_{2}(\theta,\Theta)\equiv T_{2}(\theta|\{(-1)^{n}\Theta\})$. The
local hamiltonian containing both the two-body and the three-body
interaction is then defined by \[
H=\frac{\pi}{\gamma}\frac{d}{d\theta}\log T_{2}(\theta)|_{\theta=\Theta}\]
 The inhomogeneity introduces the interaction between the left-going
frame and the right-going one.

Conveniently, we introduce an auxiliary transfer matrix which is defined
in a similar manner to $T_{2}$, replacing the spin 1 $\times$ spin
1 $R$ by the spin 1 $\times$ spin $\frac{1}{2}$ $R$ matrix. A
twist $\omega$ at the boundary can also easily be incorporated. Periodic
boundary conditions correspond to $\omega=0$, antiperiodic ones to
$\omega=\pi/2$. The explicit eigenvalues of $T_{1}(\theta)$ and
$T_{2}(\theta)$ then read \begin{eqnarray*}
T_{1}(\theta,\Theta) & = & {\textrm{e}}^{i\omega}\rho(\theta-i\pi)\frac{Q(\theta+i\pi)}{Q(\theta)}+{\textrm{e}}^{-i\omega}\rho(\theta+i\pi)\frac{Q(\theta-i\pi)}{Q(\theta)}\\
T_{2}(\theta,\Theta) & = & {\textrm{e}}^{2i\omega}\rho(\theta-i\frac{\pi}{2})\rho(\theta-i\frac{3\pi}{2})\frac{Q(\theta+i\frac{3\pi}{2})}{Q(\theta-i\frac{\pi}{2})}+\rho(\theta-i\frac{\pi}{2})\rho(\theta+i\frac{\pi}{2})\frac{Q(\theta+i\frac{3\pi}{2})Q(\theta-i\frac{3\pi}{2})}{Q(\theta+i\frac{\pi}{2})Q(\theta-i\frac{\pi}{2})}\\
 & + & \textrm{e}^{-2i\omega}\rho(\theta+i\frac{\pi}{2})\rho(\theta+i\frac{3\pi}{2})\frac{Q(\theta-i\frac{3\pi}{2})}{Q(\theta+i\frac{\pi}{2})}\end{eqnarray*}
where\[
\rho(z)=\sinh^{N/2}\frac{\gamma}{\pi}(z-\Theta)\sinh^{N/2}\frac{\gamma}{\pi}(z+\Theta)\]
 The important function $Q$ is given by the Bethe ansatz roots $\theta_{j},(j=1,\cdots,M)$;
$Q(z)=\prod_{j=1}^{M}\sinh\frac{\gamma}{\pi}(z-\theta_{j})$.

Bethe Ansatz for the homogeneous case was written in \cite{bethe-ansatz}.
For the alternating inhomogeneous case it can be generalized easily,
resulting in\[
\left(\frac{\sinh\frac{\gamma}{\pi}(\theta_{j}+\Theta+i\pi)\sinh\frac{\gamma}{\pi}(\theta_{j}-\Theta+i\pi)}{\sinh\frac{\gamma}{\pi}(\theta_{j}+\Theta-i\pi)\sinh\frac{\gamma}{\pi}(\theta_{j}-\Theta-i\pi)}\right)^{N/2}=-e^{2i\omega}\prod_{k=1}^{M}\frac{\sinh\frac{\gamma}{\pi}(\theta_{j}-\theta_{k}+i\pi)}{\sinh\frac{\gamma}{\pi}(\theta_{j}-\theta_{k}-i\pi)}\]

The number of roots $M$ is related to the (conserved) total 3-rd
component of the spin $S_{z}$ of the chain and the number of sites
$N$ by\[
S_{z}=N-M\]

The Bethe equations have a periodicity for $\theta\to\theta+\frac{i\pi^{2}}{\gamma}=\theta+i\pi(p+2)$.
It is convenient to consider as fundamental strip $-\frac{\pi(p+2)}{2}<\mathrm{Im}\theta\leq\frac{\pi(p+2)}{2}$.

The vacuum state is realized by a maximal set of (quasi)-two-strings
(see Fig. 1)\[
\theta_{j}=\mathrm{Re}\theta_{j}\pm i\left(\frac{\pi}{2}+\epsilon_{j}\right)\]
where the deviations $\epsilon_{j}>0$ from the pure two-string behavior
are assumed to be small and limited by a positive parameter $\epsilon$:
$\epsilon_{j}\leq\epsilon$. The distribution of these roots is denser
in the regions near $\pm\Theta$, where also the $\epsilon_{j}$ are
minimal. We exemplify this in some plots, at $p=\frac{1}{2}$ (attractive)
and $p=\frac{3}{2}$ (repulsive) in appendix C.

The excited states are usually characterized by roots located in positions
different than those of the vacuum, as well as some holes. The following
classification of the Bethe ansatz roots, other than 2-strings, is
possible:

\begin{enumerate}
\item inner roots : $|\mathrm{Im}\theta_{j}|<\frac{\pi}{2},\quad j=1,\cdots,M_{I}$
\item close roots: $\frac{\pi}{2}<|\mathrm{Im}\theta_{j}|<\frac{3\pi}{2},\quad j=1,\cdots,M_{C}$
\item wide roots: $\frac{3\pi}{2}<|\mathrm{Im}\theta_{j}|<\frac{\pi^{2}}{2\gamma},\quad j=1,\cdots,M_{W}$
\item self conjugate roots : $|\mathrm{Im}\theta_{j}|=\frac{\pi^{2}}{2\gamma},\quad j=1,\cdots,M_{p}$. 
\end{enumerate}
\textcolor{black}{Besides these roots, the real zeroes of the transfer
matrices $T_{1}$ and $T_{2}$ also play an important role in the
description of the excited states. The zeros of transfer matrix $T_{2}(\theta)$
are called} \textcolor{black}{\emph{holes}} \textcolor{black}{because
they are in 1:1 correspondence with actual holes in the quasi 2-string
distribution.}

Once the Bethe roots are calculated as solutions to the Bethe equations,
the formula \[
e^{i\frac{a}{2}(E\pm P)}=e^{\pm i\omega}\prod_{j=1}^{M}\frac{\sinh\frac{\gamma}{\pi}(i\pi\pm\theta_{j}-\Theta)}{\sinh\frac{\gamma}{\pi}(i\pi\mp\theta_{j}-\Theta)}\]
yields the energy level and total momentum of the corresponding state.
Higher integrals of motion can also be expressed in a similar way.

We remark that the generalization of the above formulation to general
values of spin is immediate.

\section{The NLIE for the inhomogeneous spin 1 XXZ chain\label{sec:NLIE}}

NLIE on the lattice for a homogeneous spin 1 XXZ chain with periodic
boundary conditions has been written in \cite{suzuki}, to which we
refer for notations and presentation of the deduction. The generalization
to the alternating inhomogeneous lattice model is quite straightforward
and leads, along similar lines, to the following coupled NLIE\begin{align}
\ln b(\theta) & =C_{b}+iD_{b}(\theta+i\epsilon)+ig_{1}(\theta)+ig_{b}(\theta)+(G*\ln B)(\theta)-(G^{[2\epsilon]}*\ln\bar{B})(\theta)\nonumber \\
 & +(K^{[-\frac{\pi}{2}+\epsilon]}*\ln Y)(\theta)\label{NLIElatticeb}\\
\ln y(\theta) & =C_{y}+g_{y}(\theta)+(K^{[\frac{\pi}{2}-\epsilon]}*\ln B)(\theta)+(K^{[-\frac{\pi}{2}+\epsilon]}*\ln\bar{B})(\theta)\nonumber \end{align}
 where $B(\theta)=1+b(\theta)\,,\, Y(\theta)=1+y(\theta)$ and for
any function $f$ and $g$\begin{equation}
(f*g)(x)\equiv\int\limits _{-\infty}^{+\infty}dy\, f(x-y)\, g(y)\quad,\quad f^{[\pm\eta]}(x)=f(x\pm i\eta)\label{*}\end{equation}
The kernel functions $G$ and $K$ read \[
G(\theta)=\int\limits _{-\infty}^{\infty}\frac{dq}{2\pi}\,\, e^{iq\theta}\,\,\frac{\sinh\frac{\pi(p-1)q}{2}}{2\sinh\frac{\pi pq}{2}\cosh\frac{\pi q}{2}}\qquad K(\theta)=\frac{1}{2\pi\cosh(\theta)}.\]
 We also introduce odd primitives of kernel functions, \[
\chi(\theta)=2\pi\int\limits _{0}^{\theta}dx\,\, G(x)\qquad\chi_{K}(\theta)=2\pi\int\limits _{0}^{\theta}dx\,\, K(x)\]
that are important in writing the source terms containing information
on the excitations:\begin{eqnarray*}
g_{b}(\theta) & = & \sum_{j=1}^{N_{H}}\chi(\theta-h_{j})-2\sum_{j=1}^{N_{H}^{S}}\chi(\theta-h_{j}^{S})-\sum_{j=1}^{M_{S}}\left(\chi(\theta-s_{j})+\chi(\theta-\bar{s}_{j})\right)\\
 & - & \sum_{j=1}^{M_{C}}\chi(\theta-c_{j})-\sum_{j=1}^{M_{W}}\chi_{II}(\theta-w_{j})-\sum_{j=1}^{M_{sc}}\chi_{II}(\theta-w_{sc}^{(j)}),\\
g_{1}(\theta) & = & \sum_{j=1}^{N_{1}}\chi_{K}(\theta-h_{j}^{(1)}),\\
g_{y}(\theta) & = & \lim_{\eta\rightarrow0^{+}}\tilde{g}_{y}\left(\theta+i\frac{\pi}{2}-i\eta\right)\\
\tilde{g}_{y}(\theta) & = & i\left\{ \sum_{j=1}^{N_{H}}\chi_{K}(\theta-h_{j})-2\sum_{j=1}^{N_{H}^{S}}\chi_{K}(\theta-h_{j}^{S})-\sum_{j=1}^{M_{S}}\left(\chi_{K}(\theta-s_{j})+\chi_{K}(\theta-\bar{s}_{j})\right)\right.\\
 & - & \left.\sum_{j=1}^{M_{C}}\chi_{K}(\theta-c_{j})\right\} \end{eqnarray*}
 where the second determination function, for $p>1$\[
\chi_{II}(\theta)=-i\left[\ln\sinh\left(\frac{\theta}{p}\right)-\ln\sinh\left(\frac{\theta+\mathrm{Sign}(\mathrm{Im}\theta)i\pi}{p}\right)\right]\]
 is defined by $\chi(x)+\chi(x-i\pi{\textrm{sgn }}(x))$.

The driving term bulk contribution exists only in the equation for
$\ln b(\theta)$ and it reads \[
D_{b}(\theta)=N\arctan\frac{\sinh\theta}{\cosh\Theta}\]

\textcolor{black}{The constants $C_{b}$ and $C_{y}$ are determined
by the asymptotic behavior of both sides of the NLIE}

\textcolor{black}{\begin{equation}
C_{b}=i\hat{C_{b}}\qquad\hat{C_{b}}=\pi\delta_{b}+\alpha\qquad\alpha=\omega\left(1+\frac{2}{p}\right)+\chi_{\infty}\,(N_{-}-N_{+}).\end{equation}
}

\textcolor{black}{\begin{equation}
\delta_{b}\in\{0,1\}\quad\hbox{and}\quad\chi_{\infty}=\chi(+\infty)=\frac{\pi}{2}\left(1-\frac{1}{p}\right).\end{equation}
}

\textcolor{black}{\begin{equation}
N_{\pm}=\left\lfloor 3\left(\frac{S_{z}}{p+2}\mp\frac{\omega}{\pi}\right)\right\rfloor -\left\lfloor \frac{S_{z}}{p+2}\mp\frac{\omega}{\pi}\right\rfloor \end{equation}
}where $\lfloor x\rfloor$ stands for the integer part of x and\begin{equation}
C_{y}=i\pi\delta_{y}+i\pi(S+M_{W}+M_{sc})\qquad\delta_{y}=N_{-}\,\,\hbox{mod}\,\,2\end{equation}

We need one more equation for the determination of wide and self-conjugated
roots. For the cases $\pi<\hbox{Im}\theta\leq\frac{\pi}{2}(p+1)$:

\begin{equation}
\ln\tilde{a}(\theta)=C_{\tilde{a}}+ig_{\tilde{a}}(\theta)+(G_{II}^{[-\epsilon]}*\ln B)(\theta)-(G_{II}^{[\epsilon]}*\ln\bar{B})(\theta),\end{equation}
 where \begin{eqnarray}
g_{\tilde{a}}(\theta) & = & \sum_{j=1}^{N_{H}}\chi_{II}(\theta-h_{j})-2\sum_{j=1}^{N_{H}^{S}}\chi_{II}(\theta-h_{j}^{S})-\sum_{j=1}^{M_{S}}\left(\chi_{II}(\theta-s_{j})+\chi_{II}(\theta-\bar{s}_{j})\right)\nonumber \\
 & - & \sum_{j=1}^{M_{C}}\chi_{II}(\theta-c_{j})-\sum_{j=1}^{M_{W}}\left[\chi_{II}(\theta-w_{j})\right]_{II}-\sum_{j=1}^{M_{sc}}\left[\chi_{II}(\theta-w_{sc}^{(j)})\right]_{II},\label{Dah}\end{eqnarray}
\begin{equation}
C_{\tilde{a}}=i\left\{ 2\omega\left(1+\frac{2}{p}\right)+\frac{\pi}{p}(N_{+}-N_{-})\right\} .\end{equation}

and the {}``second-second'' determination is defined as in \cite{FRT4}.

The parameters $h_{j},s_{j},c_{j},...$ in the source terms, i.e.
the positions of holes, close, wide roots, etc... can be determined
recalling that all these objects annulate the functions $B,1+\tilde{a},Y$,
etc... This leads to the following \emph{quantization conditions}

\begin{itemize}
\item For holes: \begin{equation}
\frac{1}{i}\,\ln b(h_{j}-i\epsilon)=2\pi\, I_{h_{j}}\qquad j=1,...,N_{H}.\label{eq:holes}\end{equation}

\item A bit formally for special roots: \begin{equation}
\frac{1}{i}\,\ln b(s_{j}-i\epsilon)=2\pi\, I_{s_{j}}\qquad j=1,...,M_{S}.\label{eq:specials}\end{equation}
 
\item For close roots (only for the upper part of the close pair): \begin{equation}
\frac{1}{i}\,\ln b(c_{j}^{\uparrow}-i\epsilon)=2\pi\, I_{c_{j}^{\uparrow}}\qquad j=1,...,M_{C}/2.\label{eq:close}\end{equation}
 
\item For wide roots: \begin{equation}
\frac{1}{i}\,\ln\tilde{a}(w_{j}^{\uparrow})=2\pi\, I_{w_{j}^{\uparrow}}\qquad j=1,...,M_{W}/2.\label{eq:wide}\end{equation}
 
\item For self-conjugated roots: \begin{equation}
\frac{1}{i}\,\ln\tilde{a}(w_{sc}^{\uparrow(j)})=2\pi\, I_{w_{sc}^{\uparrow(j)}}\qquad j=1,...,M_{sc}.\label{eq:self}\end{equation}
 
\end{itemize}
So far we have determined only the upper part of the complex pairs,
but the other parts can be determined by simple complex conjugation. 

\begin{itemize}
\item For zeroes of $T_{1}(\theta)$: \begin{equation}
\frac{1}{i}\,\ln y_{1}\left(h_{j}^{(1)}-i\frac{\pi}{2}\right)=2\pi\, I_{h_{j}^{(1)}}\qquad j=1,...,N_{1}.\label{eq:t1}\end{equation}

\end{itemize}
All the above quantum numbers $I_{\alpha_{j}}$'s are half integers.
A state is then identified by a choice of the quantum numbers $(I_{h_{j}},I_{c_{j}},...)$.
The NLIE itself can impose constraints on the allowed values some
of these quantum numbers.

A comment must be done to explain the emergence of \emph{special}
objects (holes and/or roots) $h_{j}^{S}$ and $s_{j},\bar{s}_{j}$
in the equations above. Normally the function $Z(\theta)=\mathrm{Im\ln b(\theta)}$
is a monotonically increasing function of $\theta$. It may happen,
however, that this monotonicity is locally violated. If it happens
that a root or hole $\theta_{j}$ has $Z(\theta_{j})<0$, then we
say that it is \emph{special}. In this case the term $\ln B=\ln(1+b)$
inside the convolution terms goes off-branch, so that $\ln B=\log B+2\pi i$
(where we use the notation log to indicate the fundamental branch
of the logarithm). The $2\pi$ jump is responsible of the emergence
of the new terms $\chi(\theta-s)+\chi(\theta-\bar{s})$ (or $2\chi(\theta-h^{S})$
for holes) in the source part of the equation. Smoothly varying parameters
in the NLIE, if an object that was not a special becomes now special,
it is compelled to be accompanied by two new emerging holes. So the
number $N_{eff}=N_{H}-2N_{s}$ is conserved by smoothly varying parameters.

A careful analysis of the $\theta\to\pm\infty$ asymptotics of the
NLIE shows that the numbers $N_{H},M_{C},M_{W},...$ etc... are not
arbitrary, but connected to the total 3-rd component of spin $S_{z}$
through a set of Diophantine equations, the so called \emph{counting
equations}\begin{equation}
N_{H}-2N_{S}=M_{C}+2(S_{z}+M_{W}+M_{sc})-N_{+}-N_{-},\label{ceNH}\end{equation}
\begin{equation}
N_{1}-2N_{R}^{S}=S_{z}+M_{W}+M_{sc}+M_{C}^{(2)}-M_{R}-\left\lfloor \frac{S_{z}}{p+2}-\frac{\omega}{\pi}+\frac{1}{2}\right\rfloor -\left\lfloor \frac{S_{z}}{p+2}+\frac{\omega}{\pi}+\frac{1}{2}\right\rfloor .\label{ceN1}\end{equation}
where $M_{C}^{(2)}$ is the number of those \char`\"{}close roots\char`\"{}
whose imaginary part are either in $\{\pi/2,\pi\}$ or $\{-\pi,-\pi/2\}$.
There is an additional parity constraint on $N_{1}$\begin{equation}
N_{1}+\left\lfloor \frac{S_{z}}{p+2}+\frac{\omega}{\pi}+\frac{1}{2}\right\rfloor -\left\lfloor \frac{S_{z}}{p+2}-\frac{\omega}{\pi}+\frac{1}{2}\right\rfloor =\hbox{even}.\end{equation}

We are interested in the \emph{scaling limit} $N\to\infty$, $a\to0$,
while the circumference of the cylinder $L=Na$ remains finite. This
limit defines the theory on a continuum cylinder of finite spatial
size (circumference) $L$. Bethe equations become infinite in number
in this limit. They must be substituted by some sort of {}``density
of roots'' method. However the traditional density approach, based
on linear integral equations, is valid only in the large $L$ limit
and does not take into account finite size effects. The NLIE approach,
instead, is able to fully control the physics on the continuum at
any size $L$.

The lattice NLIE (\ref{NLIElatticeb}) has a sensible scaling limit
only if we fine tune the inhomogeneity parameter $\Theta$ to go to
infinity as $\ln N$ in a very precise way\[
\Theta\sim\ln\frac{2N}{\mathcal{M}L}\]
thus introducing a \emph{mass scale} $\mathcal{M}$. The only variation
in the NLIE is in the $D_{b}$ term that becomes simply\[
D_{b}(\theta)=\mathcal{M}L\sinh\theta\]
The asymptotic particle interpretation of this term will be clear
in the IR analysis of next section. We are in presence of a genuine
renormalization procedure, leading to a renormalized quantum field
theory on the continuum. In the following it is convenient to introduce
the dimensionless scale parameter $\ell=\mathcal{M}L$. The UV physics
of the model is reproduced for $\ell\to0$ (negligible mass scale),
while the IR one corresponds to the {}``large volume'' $\ell\to\infty$
regime.

The counting equations on the continuum are slightly simplified with
respect to the lattice. They read

\begin{equation}
N_{H}-2N_{S}=M_{C}+2(S+M_{W}+M_{sc}),\label{eq:countingH}\end{equation}
\begin{equation}
N_{1}-2N_{R}^{S}=S+M_{W}+M_{sc}+M_{C}^{(2)}-M_{R}-\delta_{y},\label{eq:countingN1}\end{equation}
 where $S$ is now the topological charge of the continuum theory,
connected to the lattice spin $S_{z}$ is: \begin{equation}
2S=2S_{z}-N_{+}-N_{-}.\label{31}\end{equation}
The terms with integer parts on the rhs. of (\ref{ceN1}) present
on the lattice and coming from the value of Im$B(\pm\infty)$ become
zero in the continuum, where $B(\pm\infty)=1$. The same happens in
the additional parity constraint for $N_{1}$ that now simply becomes
the requirement that it must be even: $N_{1}\in2\mathbb{Z}$.

The NLIE can be solved numerically with an iterative procedure for
a given set of quantum numbers $I_{k}$. Once the position of the
sources and the functions $\ln b$, $\ln y$ have been calculated,
the total energy and momentum of a state can be calculated through
the formulae

\begin{eqnarray}
E & = & \mathcal{M}\left(\sum_{j=1}^{N_{H}}\,\cosh(h_{j})-2\sum_{j=1}^{N_{H}^{S}}\,\cosh(h_{j}^{S})-\sum_{j=1}^{M_{S}}\,\left\{ \cosh(s_{j})+\cosh(\bar{s_{j}})\right\} \right.\\
 & - & \left.\sum_{j=1}^{M_{C}}\,\cosh(c_{j})+\frac{i}{2\pi}\int\limits _{-\infty}^{+\infty}d\theta\sinh(\theta+i\epsilon)\,\ln B(\theta)-\frac{i}{2\pi}\int\limits _{-\infty}^{+\infty}d\theta\sinh(\theta-i\epsilon)\,\ln\bar{B}(\theta)\right)\nonumber \end{eqnarray}

\begin{eqnarray}
P & = & \mathcal{M}\left(\sum_{j=1}^{N_{H}}\,\sinh(h_{j})-2\sum_{j=1}^{N_{H}^{S}}\,\sinh(h_{j}^{S})-\sum_{j=1}^{M_{S}}\,\left\{ \sinh(s_{j})+\sinh(\bar{s_{j}})\right\} \right.\\
 & - & \left.\sum_{j=1}^{M_{C}}\,\sinh(c_{j})+\frac{i}{2\pi}\int\limits _{-\infty}^{+\infty}d\theta\cosh(\theta+i\epsilon)\,\ln B(\theta)-\frac{i}{2\pi}\int\limits _{-\infty}^{+\infty}d\theta\cosh(\theta-i\epsilon)\,\ln\bar{B}(\theta)\right)\nonumber \end{eqnarray}
Notice that wide and self-conjugate root contributions drop from this
expressions because $\sinh_{II}\theta=\cosh_{II}\theta=0$ in the
repulsive regime $p>1$ we are considering here.

\section{Large scale analysis and Particle Theory\label{sec:IR}}

It is interesting to verify which scattering data can be reproduced
by the large $\ell$ limit of this construction. In the $\ell\rightarrow\infty$
limit the convolution integrals involving $B$ and $\bar{B}$ in both
equations are exponentially depressed, as can be easily seen by replacing
the leading large $\ell$ contribution that goes as $\ell\sinh\theta$
into the convolution itself. However, this same reasoning cannot be
applied to the other convolution integrals involving $Y$, as the
function $y$ has no exponentially leading term. As a result, the
$\ell\to\infty$ asymptotics of the NLIE looks like\begin{align}
-i\ln b(\theta) & =\hat{C}_{b}+\ell\sinh\theta+g_{1}(\theta)+g_{b}(\theta)-i(K^{[-\frac{\pi}{2}+\epsilon]}*\ln Y)(\theta)+O(e^{-\ell})\label{eq:IR-NLIE1}\\
\ln y(\theta) & =C_{y}+g_{y}(\theta)+O(e^{-\ell})\label{eq:IR-NLIE2}\end{align}
The dominating term of the first equation for large $\ell$ is $\ell\sinh\theta$,
thus showing monotonicity of the real part of the function $-i\ln b$,
which in turn implies that for large $\ell$ there are no special
objects: $N_{s}=0$.

The second equation is now purely algebraic and can be solved for
$y$ and the result can be plugged into the first equation. Defining\[
\mathcal{K}(\theta)\equiv(K^{[-\frac{\pi}{2}]}*\ln(1+e^{C_{y}+g_{y}}))(\theta)\]
we have, putting $\omega=0$ and letting $\delta_{b}$ take both its
values 0 or 1\begin{equation}
-i\ln b(\theta)=\ell\sinh\theta+g_{b}(\theta)-i\mathcal{K}(\theta)+\delta_{b}\pi\label{eq:IR-NLIE}\end{equation}

The energy and momentum expressions also simplify in the same way\begin{eqnarray}
E & = & \mathcal{M}\,\sum_{j=1}^{N_{H}}\,\cosh(h_{j})-\mathcal{M}\,\sum_{j=1}^{M_{C}}\,\cosh(c_{j})+O(e^{-\ell})\end{eqnarray}

\begin{eqnarray}
P & = & \mathcal{M}\,\sum_{j=1}^{N_{H}}\,\sinh(h_{j})-\mathcal{M}\,\sum_{j=1}^{M_{C}}\,\sinh(c_{j})+O(e^{-\ell})\end{eqnarray}

Let us first examine the case of $N_{H}$ holes without any complex
pair (other than, of course, the quasi 2-strings of the vacuum sea,
that do not contribute here, being confined in the dropping convolution
integrals). The energy and momentum are given by\[
E=\mathcal{M}\,\sum_{j=1}^{N_{H}}\,\cosh(h_{j})\quad,\quad P=\mathcal{M}\,\sum_{j=1}^{N_{H}}\,\sinh(h_{j})\]
We can interpret this result by saying that holes represent a system
of asymptotic particles of mass $\mathcal{M}$ and rapidity $h_{j}$. 

The quantization of particle rapidities for large volume in a factorized
scattering theory is given by the formula \begin{equation}
e^{i\mathcal{M}\ell\sinh{\theta_{k}}}\Lambda(\theta_{k}|\{\theta_{i}\})=1.\label{QL}\end{equation}
where $\Lambda(\theta|\{\theta_{i}\})$ is an eigenvalue of the \emph{n-particle
color transfer matrix} defined by the products of two-body S-matrices\[
T(\theta|\{\theta_{i}\})=\prod_{i=1}^{n}S(\theta-\theta_{i})\]

A comparison of (\ref{QL}) with the exponentiated form of (\ref{eq:IR-NLIE})
and recalling the quantization conditions (\ref{eq:holes},...,\ref{eq:t1})
shows that in general\[
\Lambda(\theta_{k}|\{\theta_{i}\})=e^{ig_{b}(\theta_{k})}e^{\mathcal{K}(\theta_{k})}(-1)^{\delta_{b}}\]

In the case of two particles, $\Lambda$ are simply the eigenvalues
of the two-body S-matrix. Moreover, if the S-matrix has a tensor product
form, also its eigenvalues have. The two particle states are realized
by choosing $N_{H}=2$. The two hole positions are denoted $h_{1},h_{2}$.
We recall the well known fact that\[
e^{i\chi(\theta)}=S_{0}(\theta)\]
where $S_{0}(\theta)$ is the $S_{++}^{++}=S_{--}^{--}$ element of
the SG two-body S-matrix. The counting equation (\ref{eq:countingH})
gives a few possible choices

\begin{itemize}
\item $S=1$ and $M_{C}=M_{W}=M_{SC}=0$
\item $S=0$ and $M_{C}=2$, $M_{W}=M_{SC}=0$
\item $S=0$ and $M_{SC}=1$, $M_{C}=M_{W}=0$
\end{itemize}
Notice that $M_{W}$ and $M_{C}$ can only be even, as the complex
roots other than self-conjugate always come in pairs. From the other
counting equation (\ref{eq:countingN1}) one can see that for all
the 2-hole states $N_{1}=0$.

A simple inspection case by case of the first factor $e^{ig_{b}(\theta_{k})}$
of $\Lambda$ shows that it reproduces exactly the SG S-matrix and
the quantization of close pair imaginary parts studied in \cite{FRT2}.
We conclude that\[
\Lambda=\Lambda_{SG}e^{i\mathcal{K}}(-1)^{\delta_{b}}\]
The Ahn S-matrix will be reproduced if $e^{i\mathcal{K}}(-1)^{\delta_{b}}$
gives the four eigenvalues of the RSOS(2) S-matrix. As\[
y(\theta)=(-1)^{\delta_{y}}\tanh\left(\frac{\theta-h_{1}}{2}\right)\tanh\left(\frac{\theta-h_{2}}{2}\right)\]
for the calculation of $\mathcal{K}$ we have to distinguish the two
cases $\delta_{y}=0$ and $\delta_{y}=1$. 

In the $\delta_{y}=0$ case one gets \begin{equation}
\mathcal{K}(\theta)=i\pi\,\left\{ Q_{1}(\theta-h_{12})-Q_{2}(\theta-h_{1})-Q_{2}(\theta-h_{2})\right\} ,\end{equation}
 where $h_{12}=\frac{h_{1}+h_{2}}{2}$ and \begin{equation}
Q_{1}(\theta)=-\frac{1}{2\pi}\arctan\sinh(\theta)-\frac{i}{2\pi}\ln\cosh(\theta),\end{equation}
\begin{equation}
Q_{2}(\theta)=-\frac{\chi_{2}(\theta)}{2\pi}-\frac{i}{2\pi}\ln\cosh\left(\frac{\theta}{2}\right).\end{equation}
 where $\chi_{2}(\theta)$ is equal to $\chi(\theta)$ with $p=2$.
Important values from the point of view of the quantization equations
are\begin{equation}
{\mathcal{K}}(h_{1})={\mathcal{R}}(h_{1}-h_{2}),\end{equation}
\begin{equation}
{\mathcal{K}}(h_{2})={\mathcal{R}}(h_{2}-h_{1}),\end{equation}
 where \begin{equation}
{\mathcal{R}}(\theta)=\frac{i}{2}\chi_{2}(\theta)-\frac{i}{2}\psi_{0}(\theta),\end{equation}
\begin{equation}
\psi_{0}(\theta)=i\,\ln\frac{\sinh\frac{i\pi+\theta}{4}}{\sinh\frac{i\pi-\theta}{4}}.\end{equation}
 The following important identity yields \begin{equation}
\chi_{4}(\theta)=\frac{1}{2}\chi_{2}(\theta)+\frac{1}{2}\psi_{0}(\theta).\end{equation}
 where $\chi_{4}(\theta)$ denotes $\chi(\theta)$ calculated at $p=4$.
This gives two of the four eigenvalues of the RSOS(2) S-matrix, namely\begin{equation}
\pm e^{i\chi_{4}(\theta)}\frac{\sinh\left(\frac{i\pi+\theta}{4}\right)}{\sinh\left(\frac{i\pi-\theta}{4}\right)}\label{Q1}\end{equation}
 Along similar lines, the case $\delta_{y}=1$ gives the other two
eigenvalues\begin{equation}
\pm e^{i\chi_{4}(\theta)}\sqrt{\frac{\sinh\left(\frac{i\pi+\theta}{4}\right)}{\sinh\left(\frac{i\pi-\theta}{4}\right)}}\label{Q2}\end{equation}
thus exhausting the RSOS part of the two-body S-matrix. The complete
structure of the solitonic S-matrix of Ahn \cite{ahn} is reproduced. 

We observe that the eigenvalues (\ref{Q1},\ref{Q2}) of the RSOS
part of the color transfer matrix do not contain the $2^{\theta/2\pi i}$
crossing factors of Ahn's S-matrix. This is because the crossing factors
correspond to a gauge transformation of the RSOS transfer matrix,
under which the eigenvalues obtainable by finite size effects are
invariant \cite{rsos}.

The main surprise of this discussion is the fact that to take into
account all the 4 RSOS eigenvalues we had to take into account both
values $\delta_{y}=0,1$. Now, from the lattice construction, for
the states here examined, only $\delta_{y}=0$ is allowed. The case
where in the second NLIE $\delta_{y}$ is taken to be 1 cannot be
deduced from lattice construction. To say that there exists a NLIE
where $\delta_{y}=1$ is a conjecture, similar in a certain sense
to the one leading to the NLIE for odd topological charge sectors
in SG \cite{FRT3}. Allowing this extension of the NLIE does not only
reproduce the two missing eigenvalues of RSOS part of the S-matrix,
but, as we shall see in the following, also the entire R sector of
the UV modular invariant partition function. If there is a viable
lattice construction for this choice of $\delta_{y}$ or not remains
to be clarified.

\section{UV limit and kink approximation\label{sec:UV}}

Now we examine the ultraviolet limit of the states described by the
NLIE. We follow the standard approach described in detail in \cite{DdV1,ZamiC,KP}.
The position of the sources for $\ell\rightarrow0$ can remain finite
(central objects), or they can move towards the two infinities as
$\pm\ln\left(\frac{2}{\ell}\right)$ (left\emph{/}right movers). We
introduce the finite parts $\theta_{j}^{\pm,0}$ of their positions
$\theta_{j}$ by subtracting the divergent contribution: \[
\{\theta_{j}\}\rightarrow\left\{ \theta_{j}^{\pm}\pm\ln\left(\frac{2}{\ell}\right),\,\theta_{j}^{0}\right\} .\]
 We denote the number of right/left moving and central objects by
$N_{H}^{\pm,0},N_{S}^{\pm,0},M_{C}^{\pm,0},\dots$ etc. For later
convenience we introduce the right/left moving and central spin given
by \begin{equation}
S^{\pm,0}=\frac{1}{2}(N_{H}^{\pm,0}-2N_{S}^{\pm,0}-M_{C}^{\pm,0}-2M_{W}^{\pm,0}-2M_{sc}^{\pm,0}).\end{equation}
 In the UV limit the NLIE splits into three separate equations corresponding
to the three asymptotic regions. This is why for all the auxiliary
functions of the NLIE (\ref{NLIElatticeb}) we define the so called
\emph{kink functions} as \begin{equation}
F_{\pm}(\theta)=\lim_{\ell\rightarrow0}F\left(\theta\pm\ln\frac{2}{\ell}\right).\qquad F\in\{\log b,\,\log y,\log\tilde{a},\dots\}.\end{equation}
 In the UV limit these kink functions satisfy the so called \emph{kink
equations} and the energy and momentum can be expressed by them. Performing
the above \emph{kink limit} on our NLIE the kink equations take the
form \begin{eqnarray}
\log b_{\pm}(\theta) & = & i\, C_{b\pm}^{(i)}\pm i\, e^{\pm(\theta+i\epsilon)}+i\, g_{1\pm}^{(0)}(\theta+i\epsilon)+i\, g_{b\pm}^{(0)}(\theta+i\,\epsilon)\nonumber \\
 & + & (G*\ln B_{\pm})(\theta)-(G^{+2\epsilon}*\ln\bar{B}_{\pm})(\theta)+(K^{-\frac{\pi}{2}+\epsilon}*\ln Y_{\pm})(\theta),\label{lnbk}\end{eqnarray}
\begin{equation}
\log y_{\pm}\left(\theta-i\frac{\pi}{2}\right)=\tilde{C}_{y\pm}+\tilde{g}_{y\pm}^{(0)}(\theta)+(K^{-\epsilon}*\ln B_{\pm})(\theta)-(K^{+\epsilon}*\ln\bar{B}_{\pm})(\theta),\end{equation}
\begin{equation}
\log y_{\pm}(\theta)=\lim_{\eta\rightarrow{\frac{\pi}{2}}^{-}}\log y_{\pm}\left(\theta-i\frac{\pi}{2}+i\eta\right).\label{lny1pmdef}\end{equation}
\begin{equation}
\ln{\tilde{a}}_{\pm}(\theta)=C_{\tilde{a}}+iC_{\tilde{a}}^{\pm}+i\,\tilde{g}_{\tilde{a}\pm}^{(0)}(\theta)+(G_{II}^{-\epsilon}*\ln B_{\pm})(\theta)-(G_{II}^{+\epsilon}*\ln\bar{B}_{\pm})(\theta),\label{lastkinkeq}\end{equation}
 where in (\ref{lnbk}) $i\in\{1,2\}$ depending on the value of the
imaginary part of $\theta$. If $0<\mathrm{Im}\theta<\pi/2$, then
$i=1$, if $\pi/2<\mathrm{Im}\theta<\pi$, then $i=2$. The bulky
expressions of the constants and the concrete form of the source functions
are collected in appendix A. The finite part of the right and left
moving objects can be obtained from the kink functions by imposing
quantization conditions very similarly to (\ref{eq:holes},...,\ref{eq:t1}).

After some usual manipulations \cite{DdV1,ZamiC,KP} it turns out
that in the UV limit the energy and momentum can be expressed by a
sum of dilogarithm functions with the $\theta\rightarrow\pm\infty$
limiting values of the kink functions in their argument. One of these
two limiting values is trivial, namely: \begin{equation}
b_{\pm}(\pm\infty)=\bar{b}_{\pm}(\pm\infty)=0,\end{equation}
\begin{equation}
y_{\pm}(\pm\infty)=(-1)^{\delta_{y}}.\end{equation}
 The other limiting values satisfy a nontrivial coupled nonlinear
algebraic equations, the so-called plateau equations \cite{DdV1}:
\begin{eqnarray}
\log b_{\pm}(\mp\infty) & = & i\hat{C}_{b}\pm2i\chi_{\infty}(S-2S^{\pm})\pm2\pi i\,\hat{l}_{W}^{\pm}\pm i\pi\left(M_{sc}+\frac{N_{1}}{2}-N_{1}^{\pm}\right)\nonumber \\
 & + & \frac{\chi_{\infty}}{\pi}\,\left\{ \ln B_{\pm}(\mp\infty)-\ln\bar{B}_{\pm}(\mp\infty)\right\} +\frac{1}{2}\ln Y_{\pm}(\mp\infty),\label{logbplat}\end{eqnarray}
\begin{equation}
y_{\pm}(\mp\infty)=(-1)^{2S^{\pm}+\delta_{y}}\,\left(B_{\pm}(\mp\infty)\bar{B}_{\pm}(\mp\infty)\right)^{\frac{1}{2}},\label{Nrp}\end{equation}
 where \begin{equation}
\hat{l}_{W}^{\pm}=l_{W}^{\pm}-\frac{M_{W}^{\pm}}{2}-M_{sc}^{\pm},\end{equation}
 irrelevant from the point of view of the exponent of equation (\ref{logbplat}).
Now we have to solve eqs. (\ref{logbplat},\ref{Nrp}) analytically.
Motivated by the TBA formulation of the problem, we take the following
Ansatz for the solutions of the plateau equations (\ref{logbplat},\ref{Nrp}):
\begin{equation}
b_{\pm}(\mp\infty)=e^{\pm3i\rho_{\pm}}\,2\,\cos(\rho_{\pm}),\qquad\bar{b}_{\pm}(\mp\infty)=e^{\mp3i\rho_{\pm}}\,2\,\cos(\rho_{\pm}),\label{ap1}\end{equation}
\begin{equation}
B_{\pm}(\mp\infty)=e^{\pm2i\rho_{\pm}}\,\frac{\sin\left(3\rho_{\pm}\right)}{\sin\left(\rho_{\pm}\right)},\qquad\bar{B}_{\pm}(\mp\infty)=e^{\mp2i\rho_{\pm}}\,\frac{\sin\left(3\rho_{\pm}\right)}{\sin\left(\rho_{\pm}\right)},\label{ap2}\end{equation}
\begin{equation}
y_{\pm}(\mp\infty)=\frac{\sin\left(3\rho_{\pm}\right)}{\sin\left(\rho_{\pm}\right)},\qquad Y_{\pm}(\mp\infty)=4\,{\cos(\rho_{\pm})}^{2}>0.\label{ap3}\end{equation}
 Since we need only $b_{\pm}(\mp\infty)$ and not the (extended) log
of it, we can restrict the allowed values of $\rho_{\pm}$ in the
$[0,2\pi]$ interval. The solutions of the plateau equations formally
take the form \begin{equation}
\rho_{\pm}=\pi\left(k_{\rho}^{\pm}\pm\delta_{b}\pm\frac{\alpha}{\pi}+\Delta_{\rho_{\pm}}\right)-\frac{\pi}{p+2}\left(2k_{\rho}^{\pm}\pm2\delta_{b}\pm2\frac{\alpha}{\pi}+3\Delta_{\rho_{\pm}}\right),\label{rho1}\end{equation}
 where \begin{equation}
\Delta_{\rho_{\pm}}=S-2S^{+}-N_{\rho}^{\pm},\qquad N_{\rho}^{\pm}=\left\lfloor 3\frac{\rho_{\pm}}{\pi}\right\rfloor -\left\lfloor \frac{\rho_{\pm}}{\pi}\right\rfloor ,\qquad n_{\rho}^{\pm}=\left\lfloor 2\frac{\rho_{\pm}}{\pi}\right\rfloor -\left\lfloor \frac{\rho_{\pm}}{\pi}\right\rfloor .\end{equation}
 and further constraints must be satisfied by the parity of the integers
$k_{\rho}^{\pm}$ and $N_{\rho}^{\pm}$: \begin{equation}
k_{\rho}^{\pm}=2l_{\rho}^{\pm}+M_{sc}+\frac{N_{1}}{2}-N_{1}^{\pm}-n_{\rho}^{\pm},\qquad l_{\rho}^{\pm}\in\mathbb{Z},\label{cons+}\end{equation}
\begin{equation}
N_{\rho}^{\pm}\,\,\mbox{mod}\,\,2=2S^{\pm}+\delta_{y}\,\,\mbox{mod}\,\,2\label{N-Sp}\end{equation}
 Using the dilogarithm sum rule of appendix B. and putting everything
together the conformal weights take the form \begin{equation}
\Delta^{\pm}=\frac{1}{16}\,\delta_{y}+\frac{1}{2}\left(\frac{n_{\pm}}{R}\pm\frac{S}{2}R\right)^{2}+\tilde{N}_{\pm}+J_{\pm},\label{DELTA}\end{equation}
 where \begin{equation}
n_{\pm}=-\left(\delta_{b}\pm k_{\rho}^{\pm}\pm\frac{3}{2}\Delta_{\rho_{\pm}}\right),\label{nplusminus}\end{equation}
\begin{equation}
\tilde{N}_{\pm}=\frac{\hat{N}_{\rho}^{\pm}-\delta_{y}}{8}\mp S^{\pm}n_{\pm}-\frac{3}{2}(S-S^{\pm})S^{\pm},\qquad\hat{N}_{\rho}^{\pm}=N_{\rho}^{\pm}\,\,\mbox{mod}\,\,2.\label{Npm}\end{equation}
\begin{eqnarray}
J_{\pm} & = & \pm I_{h^{\pm}}\mp2I_{h_{s}^{\pm}}\mp2I_{s^{\pm}}\mp2I_{c^{\pm\uparrow}}\mp2I_{w^{\pm\uparrow}}\mp I_{w_{sc}^{\pm\uparrow}}\pm I_{h^{(1)\pm}}\nonumber \\
 & \mp & (S^{\pm}+M_{W}^{\pm}+M_{sc}^{\pm})\,\left(\delta_{b}\pm\frac{N_{1}}{2}\mp N_{1}^{\pm}\pm M_{sc}\mp M_{sc}^{\pm}\right)\\
 & - & (M_{W}^{+}+M_{sc}^{+})\,\left(S-S^{\pm}\right)+K_{\pm}.\nonumber \end{eqnarray}
 where $K_{+}$ and $K_{-}$ are given by \begin{equation}
K_{+}=k_{W}^{+}+M_{C}^{(2)+}\,(N_{1}-N_{1}^{+})-\frac{N_{1}^{+}}{2}\left\{ \delta_{y}+2(S+M_{W}+M_{sc})+M_{C}^{(2)}-M_{C}^{(2)+}\right\} \label{Kp}\end{equation}
\begin{equation}
K_{-}=k_{W}^{-}+\frac{N_{1}^{-}}{2}\left(\delta_{y}-M_{C}^{(2)}+M_{C}^{(2)-}\right)\label{Km}\end{equation}
 where $k_{W}^{\pm}$ depend only on the actual configuration of wide
roots, and they are integers, if $M_{sc}^{\pm}$ are even, and they
are half-integers if $M_{sc}^{\pm}$ are odd.

Analyzing the formulae (\ref{DELTA}-\ref{Km}) of the UV conformal
weights, one can see that $S$ can be identified with the winding
number $m$, the parameter $\delta_{y}\in\{0,1\}$ of the NLIE distinguishes
the NS and R sectors of the theory, and depending on the state under
consideration the sum $N_{\pm}+J_{\pm}$ can be either integer or
half-integer. Furthermore it can be easily proven that there is a
relation between $n_{\pm}$ of (\ref{nplusminus}) and the winding
number $S$, namely in the NS sector \begin{equation}
n_{\pm}\in\mathbb{Z}\qquad\mbox{if}\qquad S\in2\mathbb{Z},\label{f1ns}\end{equation}
\begin{equation}
n_{\pm}\in\mathbb{Z}+\frac{1}{2}\qquad\mbox{if}\qquad S\in2\mathbb{Z}+1,\label{f2ns}\end{equation}
 while in the Ramond sector \begin{equation}
n_{\pm}\in\mathbb{Z}+\frac{1}{2}\qquad\mbox{if}\qquad S\in2\mathbb{Z},\label{f1r}\end{equation}
\begin{equation}
n_{\pm}\in\mathbb{Z}\qquad\mbox{if}\qquad S\in2\mathbb{Z}+1.\label{f2r}\end{equation}
 in full accordance with what described on the UV operator content
in section \ref{sec:SSGmodel}. Moreover it can be proven that in
the $\delta_{y}=1$ case (R sector) the sum $N_{\pm}+J_{\pm}$ is
always integer as it must be in the R sector of a $c=\frac{3}{2}$
CFT. Thus, due to the previous remark and the relations (\ref{f1ns}-\ref{f2r}),
the conformal weights (\ref{DELTA}-\ref{Km}) can be interpreted
within the framework of $c=\frac{3}{2}$ CFT as conformal weights
appearing in the modular invariant partition function of Di Francesco
et al. \cite{DSZ}.

\section{Comparison with PCFT results\label{sec:PCFT}}

In the previous section we saw that the NLIE (\ref{NLIElatticeb})
describes a modular invariant $c=\frac{3}{2}$ CFT in the UV, but
we would like to have more evidence that our coupled NLIEs describe
the SSG theory. The simplest check is the numerical comparison of
NLIE and perturbed conformal field theory for certain states. \textcolor{black}{This
comparison is really important in the R sector where the form of the
NLIE is a pure conjecture and can not be obtained from the integrable
light-cone lattice regularization of the model.}

As we already discussed in section \ref{sec:SSGmodel}, the SSG model
can be formulated as a perturbed CFT with the action on the cylinder
\begin{equation}
{\mathcal{A}}={\mathcal{A}}_{3/2}+g\,\int d^{2}z\,\,\Phi_{pert}(z,\bar{z}),\end{equation}
 where ${\mathcal{A}}_{3/2}$ is the action of the $c=\frac{3}{2}$
CFT, $g=\frac{\mu}{2\sqrt{2}\pi}$ is the coupling constant with mass
dimension $y=2(1-1/R^{2})$ with $R$ being the compactification radius
of the conformal normalized boson field and the $\Phi_{pert}(z,\bar{z})$
perturbing operator of eq.(\ref{eq:pert}) is normalized by $\langle\Phi_{pert}(z,\bar{z})\Phi_{pert}(0,0)\rangle=|z|^{-2(2-y)}$.
In the framework of conformal perturbation theory the energy levels
can be represented as a sum of integrals over the n-point functions
of the unperturbed CFT \cite{CPT}. In the case of SSG theory there
is no bulk energy contribution on account of supersymmetry, and the
perturbation series contains only even powers of the coupling $g$.
Using the exact relation between $g$ and the massgap of the model
\cite{FB}\begin{equation}
\frac{g\pi}{\sqrt{8}}\,\gamma\left(\frac{1}{2}-\frac{\beta^{2}}{32\pi}\right)=\mathcal{M}^{1-\frac{\beta^{2}}{16\pi}}\left(\frac{\pi}{4}\frac{\beta^{2}}{16\pi-\beta^{2}}\right)^{1-\frac{\beta^{2}}{16\pi}},\qquad\gamma(x)=\frac{\Gamma(x)}{\Gamma(1-x)},\end{equation}
 the perturbation series can be expressed as power series of $\ell^{2y}$,
where $\ell=\mathcal{M}L$ is the dimensionless volume. The first
correction to the energy of a state $|a\rangle$ having conformal
dimensions $\Delta_{a},\bar{\Delta}_{a}$ can be written as \begin{equation}
\frac{6L}{\pi}E_{a}(\ell)=-(\frac{3}{2}-12(\Delta_{a}+\bar{\Delta}_{a}))-C_{2}^{|a\rangle}\ell^{2y}+O(\ell),\end{equation}
 where the leading order coefficient reads as \begin{equation}
C_{2}^{|a\rangle}=\alpha\frac{I^{|a\rangle}}{\pi}\end{equation}
 where \begin{equation}
\alpha=\frac{3}{2}\cdot8^{{2}/{R^{2}}}\frac{1}{\gamma^{2}\left(\frac{1}{2}-\frac{1}{2R^{2}}\right)}\left(\frac{1}{R^{2}-1}\right)^{2-\frac{2}{R^{2}}}.\end{equation}
 contains the massgap formula and $I^{|a\rangle}$ is a simple integral
expression \begin{equation}
I^{|a\rangle}=\int\,\frac{d^{2}z}{|z|^{y}}\,\langle a|\Phi_{pert}(1,1)\Phi_{pert}(z,\bar{z})|a\rangle{}_{\mbox{conn.}}\end{equation}
 In \cite{Gabor1} the first nontrivial PCFT corrections were calculated
for various states in the NS sector and they were used to test the
TCSA near UV, and to check the NLIEs proposed to describe the three
ground states of the model. Now, having NLIE for excited states as
well, it is worthwhile to check it against PCFT predictions. Unfortunately
the comparison is quite restricted on both sides. Actually, the applicability
of PCFT, mostly in the R sector, is strongly restricted by IR divergencies.
Thus, up to two-kink states the number of IR safe states is very small.
On the other hand the NLIE can not be solved numerically with present
techniques if special objects appear \cite{DdV1}. Thus, the PCFT-NLIE
comparison can be done only for a few states in certain restricted
regions of the coupling constants.

In the NS sector, in the zero momentum two-kink states space, the
energy of the state $|1-\rangle=\frac{1}{\sqrt{2}}\left(V_{1,0}^{(0,0)}(0,0)-V_{-1,0}^{(0,0)}(0,0)\right)|0\rangle$
with conformal weights $\Delta=\bar{\Delta}=\frac{1}{2R^{2}}$ can
be calculated numerically from NLIE in the region $1<p<2$. The PCFT
correction of the state was calculated in \cite{Gabor1} and is given
by \begin{equation}
I^{|1-\rangle}=\gamma\left(\frac{1}{2}-\frac{1}{2R^{2}}\right)\left[\gamma\left(-\frac{1}{R^{2}}\right)\gamma\left(\frac{1}{2}+\frac{3}{2R^{2}}\right)-\frac{1}{2}\gamma\left(\frac{1}{2}-\frac{1}{2R^{2}}\right)\gamma\left(\frac{1}{R^{2}}\right)\right].\end{equation}
 In the language of Bethe Ansatz this state corresponds to a $S=0$
state with 2-holes, one self-conjugated root, and $\delta_{b}=\delta_{y}=0$.
We checked numerically our NLIE against PCFT at a lot of values of
the compactification radius $R$, and in every case we experienced
very good agreement. To illustrate the agreement between NLIE and
PCFT, the numerical comparison of the quantity $\varepsilon(l)=\frac{6L}{\pi}E_{a}(l)+(\frac{3}{2}-12(\Delta_{a}+\bar{\Delta}_{a}))$
coming from NLIE and PCFT at the specific $R^{2}=\frac{7}{3}$ point
can be found in table 1. One can see that the two sets of data approach
one another as the volume tends to zero.

In the space of zero-momentum two-kink states of the Ramond sector,
because of IR divergencies, the only state for which PCFT can be applied
is a charged 2-kink state. At UV, this state corresponds to $|S=1,R\rangle=R_{0,1}(0,0)|0\rangle$
with conformal weights $\Delta=\bar{\Delta}=\frac{1}{16}+\frac{1}{2}\left(\frac{R}{2}\right)^{2}$,
and the first nontrivial correction to this state is given by \begin{equation}
I^{|S=1,R\rangle}=-\frac{\pi}{4}\,\gamma\left(-\frac{1}{2}+\frac{1}{2R^{2}}\right)\,\gamma\left(\frac{1}{2}+\frac{1}{2R^{2}}\right)\,\gamma\left(-\frac{1}{R^{2}}\right).\end{equation}
 In the language of Bethe Ansatz, this state corresponds to a pure
two-hole state with $\delta_{b}=0,\delta_{y}=1$. The numerical comparison
of $\varepsilon(\ell)$ coming from NLIE and PCFT at the specific
value $R^{2}=\frac{5}{3}$ can be found in table 2. The two sets of
data converge as the volume tends to zero. 

\begin{table}
\begin{center}\begin{tabular}{|c|c|c|c|}
\hline 
 $\ell$&
 $\varepsilon(\ell)\quad\mbox{(NLIE)}$&
 $\varepsilon(\ell)\quad\mbox{(PCFT)}$&
 $\ell^{4y}$\tabularnewline
\hline
0.1 &
 0.213251889(1) &
 0.2135198827 &
 $5\cdot10^{-3}$\tabularnewline
\hline
0.05 &
 0.096639933(5) &
 0.096695005 &
 $10^{-3}$\tabularnewline
\hline
0.01 &
 0.015365337(5) &
 0.01536672004 &
 $2\cdot10^{-5}$\tabularnewline
\hline
0.005 &
 0.0069587125(5) &
 0.006959000 &
 $5\cdot10^{-6}$\tabularnewline
\hline
0.001 &
 0.001105913(5) &
 0.00110592082 &
 $10^{-7}$\tabularnewline
\hline
0.0005 &
 0.00500832(5) &
 0.005008293 &
 $3\cdot10^{-8}$\tabularnewline
\hline
0.0001 &
 0.000079592(5) &
 0.00007959153 &
 $7\cdot10^{-10}$ \tabularnewline
\hline
\end{tabular}\end{center}

\caption{{\footnotesize Numerical comparison of PCFT with NLIE for the state
$|1-\rangle$ at $R^{2}=\frac{7}{3}$. }}
\end{table}

\begin{table}
\begin{center}\begin{tabular}{|c|c|c|c|}
\hline 
 $\ell$&
 $\varepsilon(\ell)\quad\mbox{(NLIE)}$&
 $\varepsilon(\ell)\quad\mbox{(PCFT)}$&
 $\ell^{4y}$\tabularnewline
\hline
0.1 &
 0.4411699(1) &
 0.42645819 &
 $0.025$\tabularnewline
\hline
0.05 &
 0.24978747(1) &
 0.24493591 &
 $8\cdot10^{-3}$\tabularnewline
\hline
0.01 &
 0.06795829(1) &
 0.067589069 &
 $6\cdot10^{-4}$\tabularnewline
\hline
0.005 &
 0.03894153(2) &
 0.03881972 &
 $2\cdot10^{-4}$\tabularnewline
\hline
0.001 &
 0.01072140(3) &
 0.010712145 &
 $10^{-5}$\tabularnewline
\hline
0.0005 &
 0.00615559(2) &
 0.00615251 &
 $5\cdot10^{-6}$\tabularnewline
\hline
0.0001 &
 0.00016979(1) &
 0.000169776 &
 $4\cdot10^{-7}$ \tabularnewline
\hline
\end{tabular}\end{center}

\caption{{\footnotesize Numerical comparison of PCFT with NLIE for the state
$|S=1,R\rangle$ at $R^{2}=5/3$. }}
\end{table}

The very good agreement of NLIE and PCFT data give an additional strong
support that the NLIE (\ref{NLIElatticeb}) indeed describes the finite
size effects of the SSG model.

\section{Conclusions and Perspectives\label{sec:Conclusions}}

In the present paper we have presented the full excited states NLIE
for the N=1 super-sine-Gordon model, deducing it from a lattice NLIE
introduced by Suzuki for the spin 1 integrable XXZ chain \cite{suzuki}.
We have checked the consistency of this procedure by computing some
of the infrared and ultraviolet properties known for such quantum
field theory. In particular, the S-matrix proposed by C. Ahn \cite{ahn}
has been reproduced, as well as various properties of the $c=\frac{3}{2}$
CFT that is expected to describe the ultraviolet behavior of the model.
Much more work has to be done, however. First of all, contrary to
the sine-Gordon case, we lack here of a deduction of the equation
of motion from the lattice construction. This keeps the whole construction
quite conjectural (although on a very solid ground). A neat deduction
of equations of motion from the light-cone lattice construction or
alternatively from a bosonization of the alternating inhomogeneous
XXZ Hamiltonian could be a very important step forward.

We have investigated in detail only the solitonic sector of the model,
by focalizing on the repulsive regime where no breathers appear. The
structure of the supersymmetric breathers should be investigated for
at least two reasons: it completes the analysis of the asymptotic
particle spectrum of the theory and gives access to quantum group
restrictions describing superconformal minimal models perturbed by
$\Phi_{13}$.

The N=1 SSG model is just the first in a whole series of Fractional
SSG models \cite{BLeC} that can be seen as perturbations of the conformal
models introduced in \cite{DSZ}. The full series of coupled NLIE's
generalizing those presented here has been proposed, for the vacuum
state, by C. Dunning \cite{dunning}. These models are evidently mapped,
by analogy with SG and SSG, into light-cone higher vertex models like
the 44-vertex, etc... or equivalently in alternating inhomogeneous
XXZ chains of higher and higher spin. It would be interesting to check
Dunning's conjecture and extend NLIEs to the structure of excited
states. Two applications of this could be of importance: the quantum
group restriction leading to $SU(2)$-coset theories \cite{ravanini}
perturbed by $\phi_{1,3}^{(1)}$ and the large spin limit where this
series should make contact with the N=2 SSG model, whose finite size
effects are of principal importance in the formulation of superstrings
propagating in pp-wave background \cite{pp-wave}.

An even more unifying point of view has been proposed in \cite{hegedus},
where two massive coupled NLIEs encompass the whole set of $SU(2)$-cosets,
of FSSG models and their parent theory, the so-called SS-model \cite{fateev-SS}.

Finally an investigation of the SSG model with integrable boundaries,
along the lines of what done for SG in \cite{boundary} would be of
interest.

We hope to return to these issues in future.

\section*{Acknowledgments}

First of all we would like to thank Clare Dunning for her interest
in this work during its early stages and for many useful discussions.
JS would like to thank people at the University of Bologna for their
kind hospitality, and members of the University of Wuppertal for discussions.
He acknowledges the financial support by the Ministry of Education
of Japan, a Grant-in-Aid for Scientific Research 14540376. FR and
AH acknowledge the European network EUCLID (HPRN-CT-2002-00325) and
the INFN grant TO12 for partial financial support. AH also thanks
Hungarian National Science fund OTKA (under T043159 and T049495).
FR also acknowledges NATO grant PST.CLG.980424 for partial financial
support.

\section*{Appendix A}

This appendix is devoted to list the constants and source functions
appearing in the kink equations (\ref{lnbk}-\ref{lastkinkeq}). The
list of values of the constants is as follows \begin{equation}
C_{b\pm}^{(1)}=\hat{C}_{b}+2\chi_{\infty}\,(S-S^{\pm})\pm\pi\,(M_{W}-M_{W}^{\pm}+M_{sc}-M_{sc}^{\pm})+2\pi l_{W}^{\pm}\pm\frac{\pi}{2}(N_{1}-N_{1}^{\pm}),\end{equation}
\begin{equation}
C_{b+}^{(2)}=C_{b+}^{(1)}-2\pi(N_{1}-N_{1}^{+}),\qquad C_{b-}^{(2)}=C_{b-}^{(1)},\end{equation}
\begin{equation}
\tilde{C}_{y+}=i\,\pi\,\left\{ \delta_{y}+2(S+M_{W}+M_{sc})+M_{C}^{(2)}-(S^{+}+M_{W}^{+}+M_{sc}^{+}+M_{C}^{(2)+})\right\} ,\end{equation}
\begin{equation}
\tilde{C}_{y-}=i\,\pi\,\left\{ \delta_{y}-M_{C}^{(2)}+(S^{-}+M_{W}^{-}+M_{sc}^{-}+M_{C}^{(2)-})\right\} .\end{equation}
\begin{equation}
C_{\tilde{a}}=i\left\{ 2\,\omega\left(1+\frac{2}{p}\right)+\frac{\pi}{p}(N_{+}-N_{-})\right\} .\end{equation}
\begin{equation}
C_{\tilde{a}\pm}=\pm4\chi_{\infty}(S-S^{\pm})\mp2\pi(S-S^{\pm})+2\pi q_{\tilde{a}}^{\pm},\end{equation}
 The numbers $q_{\tilde{a}}^{\pm}$ and $l_{W}^{\pm}$ are integers
which appear only in the presence of wide roots, and their actual
value depends on the relative position of the wide roots. The form
of the source functions of (\ref{lnbk}-\ref{lastkinkeq}) are given
by \begin{equation}
g_{1\pm,0}^{(0)}(\theta)=\sum_{j=1}^{N_{1}^{\pm,0}}\chi_{K}(\theta-h_{j}^{(1)\pm,0}),\label{g1k}\end{equation}
\begin{eqnarray}
\tilde{g}_{y\pm,0}^{(0)}(\theta) & = & i\left\{ \sum_{j=1}^{N_{H}^{\pm,0}}\chi_{K}(\theta-h_{j}^{\pm,0})-2\,\sum_{j=1}^{N_{H}^{S\pm,0}}\chi_{K}(\theta-h_{j}^{S,\pm,0})-\sum_{j=1}^{M_{S}^{\pm,0}}\left(\,\chi_{K}(\theta-s_{j}^{\pm,0})+\chi_{K}(\theta-\bar{s}_{j}^{\pm,0})\right)\right.\nonumber \\
 & - & \left.\sum_{j=1}^{M_{C}^{\pm,0}}\chi_{K}(\theta-c_{j}^{\pm,0})\right\} .\end{eqnarray}
\begin{eqnarray}
g_{b\pm,0}^{(0)}(\theta) & = & \sum_{j=1}^{N_{H}^{\pm,0}}\chi(\theta-h_{j}^{\pm,0})-2\,\sum_{j=1}^{N_{H}^{S\pm,0}}\chi(\theta-h_{j}^{S\pm,0})-\sum_{j=1}^{M_{S}^{\pm,0}}\left(\,\chi(\theta-s_{j}^{\pm,0})+\chi(\theta-\bar{s}_{j}^{\pm,0})\right)\nonumber \\
 & - & \sum_{j=1}^{M_{C}^{\pm,0}}\chi(\theta-c_{j}^{\pm,0})-\sum_{j=1}^{M_{W}^{\pm,0}}\chi_{II}(\theta-w_{j}^{\pm,0})-\sum_{j=1}^{M_{sc}^{\pm,0}}\chi_{II}(\theta-w_{sc}^{(j)\pm,0}),\label{gbk}\end{eqnarray}
\begin{eqnarray}
\tilde{}_{\tilde{a}\pm,0}^{(0)}(\theta) & = & \sum_{j=1}^{N_{H}^{\pm,0}}\chi_{II}(\theta-h_{j}^{\pm,0})-2\,\sum_{j=1}^{N_{H}^{S\pm,0}}\chi_{II}(\theta-h_{j}^{S\pm,0})-\sum_{j=1}^{M_{S}^{\pm,0}}\left(\,\chi_{II}(\theta-s_{j}^{\pm,0})+\chi_{II}(\theta-\bar{s}_{j}^{\pm,0})\right)\nonumber \\
 & - & \sum_{j=1}^{M_{C}^{\pm,0}}\chi_{II}(\theta-c_{j}^{\pm,0})-\sum_{j=1}^{M_{W}^{\pm,0}}\left[\chi_{II}(\theta-w_{j}^{\pm,0})\right]_{II}-\sum_{j=1}^{M_{sc}^{\pm,0}}\left[\chi_{II}(\theta-w_{sc}^{(j)\pm,0})\right]_{II}.\label{Dahk}\end{eqnarray}

\section*{Appendix B}

In section \ref{sec:UV}, in the calculation of the UV conformal weights,
the following dilogarithmic sum must be calculated: \begin{equation}
S_{0}(\rho)=2\left\{ L_{+}\left[e^{3i\rho}\,2\,\cos(\rho)\right]+L_{+}\left[e^{-3i\rho}\,2\,\cos(\rho)\right]+L_{+}\left[\frac{\sin\left(3\rho\right)}{\sin\left(\rho\right)}\right]\right\} ,\label{S0}\end{equation}
 where \begin{equation}
L_{+}(x)=\frac{1}{2}\int\limits _{0}^{x}\, dy\,\left\{ \frac{\ln(1+y)}{y}-\frac{\ln y}{1+y}\right\} .\end{equation}
 The function $S_{0}(\rho)$ has the following simple properties:
\begin{equation}
S_{0}(\rho)=S_{0}(\rho+\pi),\qquad S_{0}(\rho)=S_{0}(-\rho).\end{equation}
 After some algebra one can prove that \begin{equation}
S_{0}(\rho)=\frac{2\pi^{2}}{3}+\hat{N_{\rho}}\left\{ 2\pi\left|\rho-\pi\left[\frac{\rho}{\pi}\right]-\frac{\pi}{2}\right|-\pi^{2}\right\} -i\,\hat{N_{\rho}}\pi\,\ln\left(4\,\cos^{2}\rho\right),\label{RS0}\end{equation}
 where \begin{equation}
\hat{N}_{\rho}=N_{\rho}\,\,\mbox{mod}\,\,2\qquad N_{\rho}=\left[3\frac{\rho}{\pi}\right]-\left[\frac{\rho}{\pi}\right],\end{equation}
 and we made the choice of $\ln(-1)=i\pi$.

\section*{Appendix C}

We solve the Bethe ansatz equation on the lattice directly for $N=36,\,\gamma=\frac{\pi}{2.5}$
and for various particular $\ell=2N{\textrm{e}}^{-\Theta}$ in the
ground state. This corresponds to the attractive regime and $\delta_{y}=0$.
The Bethe ansatz roots and $\Im m(\ln b)$ are plotted below. Here
we use the variable $x=\gamma\theta/\pi$.

One sees that the central regime in the graph of $\ln b$ becomes
flatter for smaller $\ell$. The slope however seems to be always
positive.

\begin{figure}

\caption{The Bethe ansatz roots (left) and the corresponding $\Im m(\ln b)$
for $N=36,\gamma=\pi/2.5,\ell=1$}

\begin{center}\includegraphics{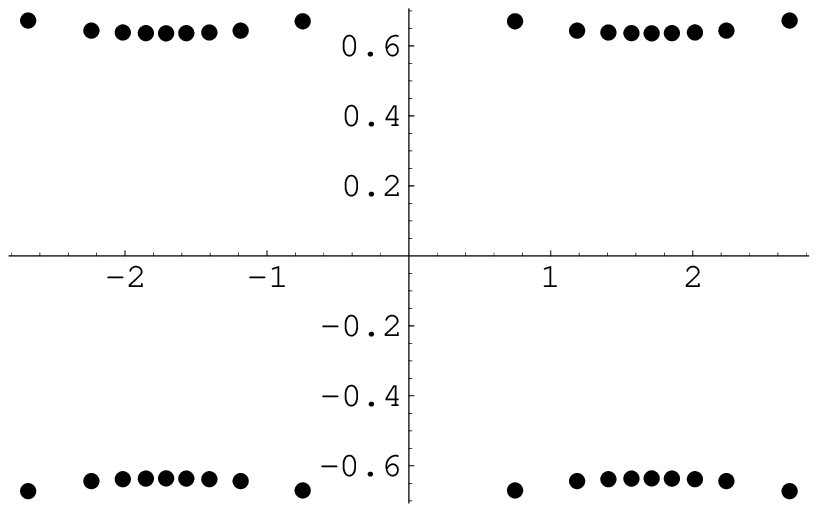}~~~~~~\includegraphics{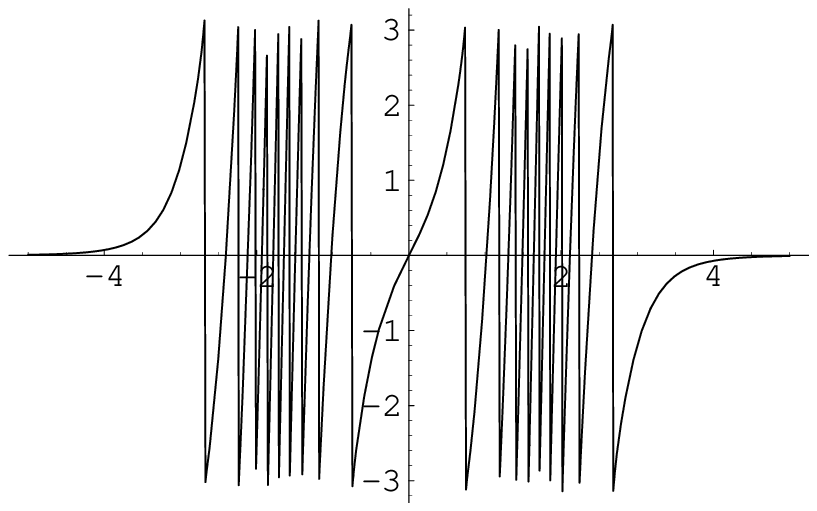}\label{N18ell1}\end{center}
\end{figure}

\begin{figure}

\caption{The Bethe ansatz roots (left) and the corresponding $\Im m(\ln b)$
for $N=36,\gamma=\pi/2.5,\ell=0.1$}

\begin{center}\includegraphics{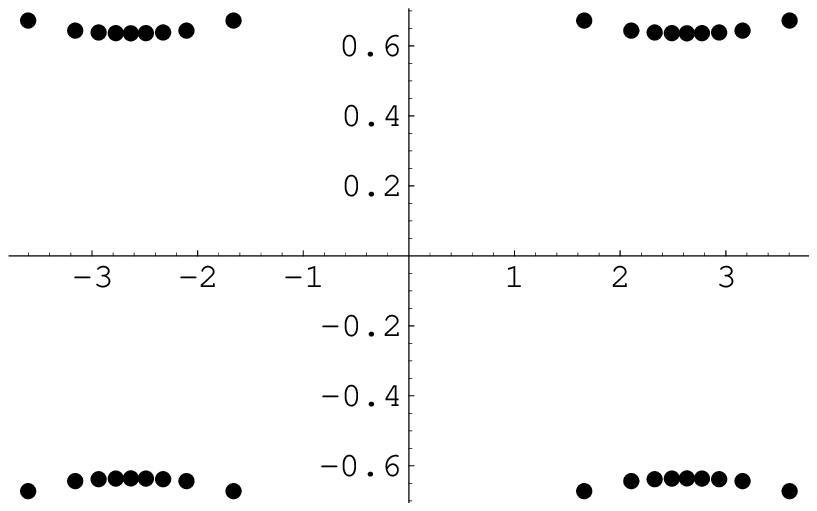} ~~~~~~\includegraphics{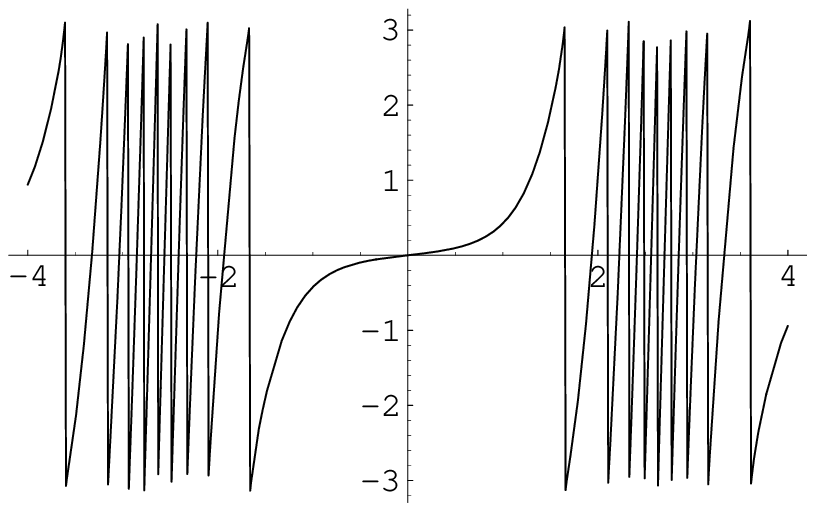}\end{center}

\label{N18ell01}
\end{figure}

\begin{figure}

\caption{The Bethe ansatz roots (left) and the corresponding $\Re(\ln b)$
for $N=128,\gamma=\pi/2.5,\ell=0.1$}

\begin{center}\includegraphics{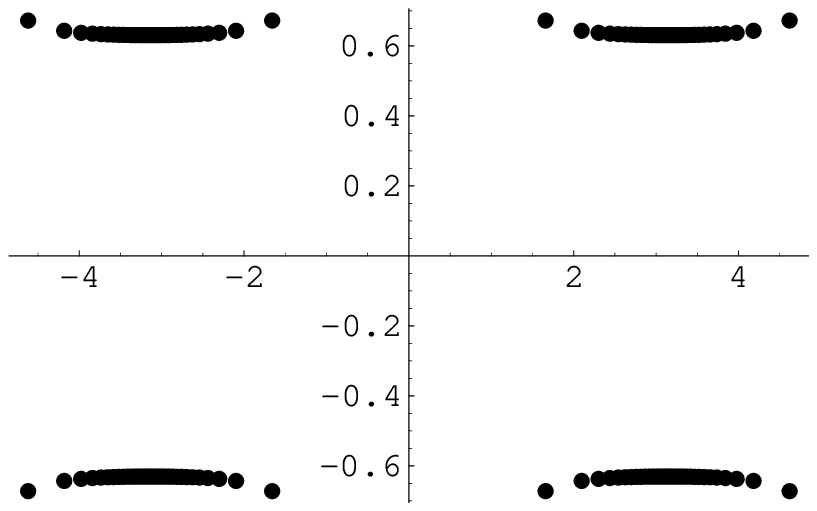}~~~~~\includegraphics{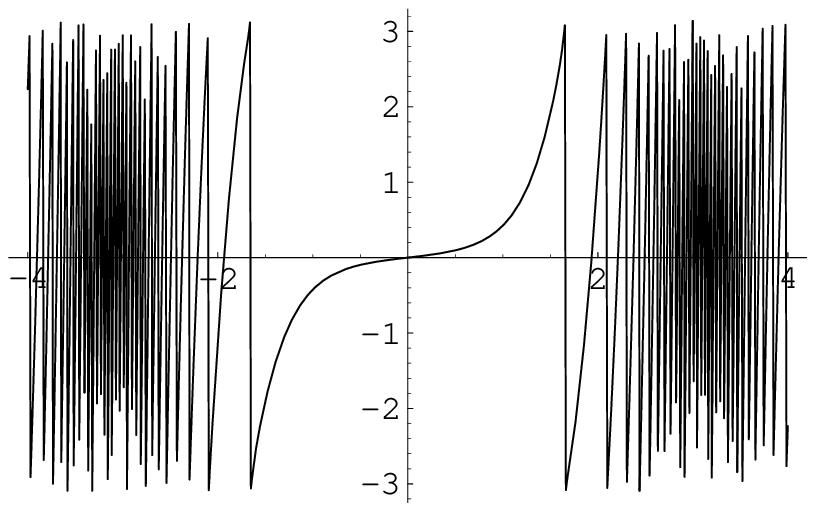}\label{WKBN18ell01}\end{center}
\end{figure}

With increase of $N$, the Newton method for finding the Bethe ansatz
roots becomes quite unstable. We thus adopt a kind of WKB method to
locate them. We plot the result for $N=128$ in Fig. \ref{WKBN18ell01}.
This again support the positivity of the slope qualitatively.

\begin{figure}

\caption{The Bethe ansatz roots (left) and the corresponding $\Im m(\ln b)$
for $N=38,\gamma=\pi/2.5,\ell=0.1$}

\begin{center}\includegraphics{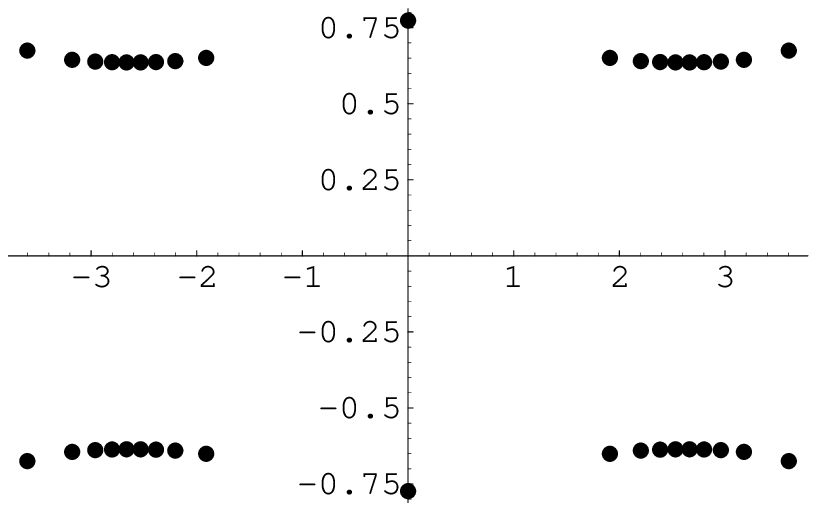}~~~~~\includegraphics{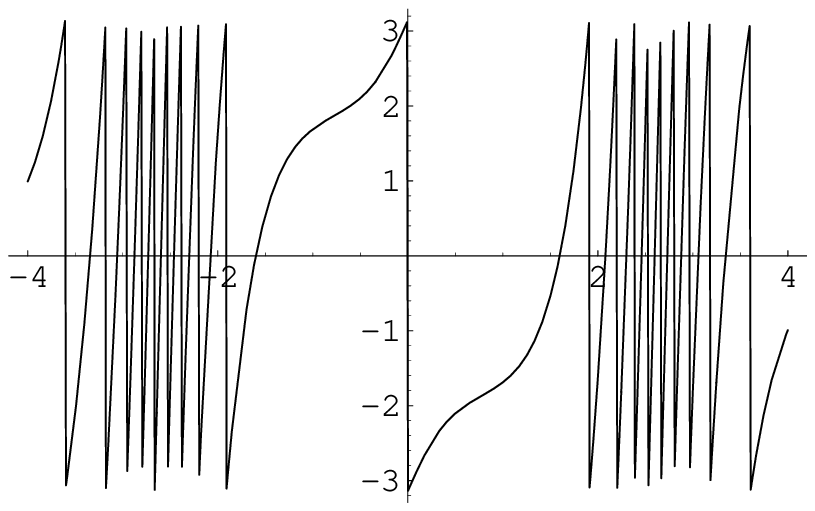}\label{N19ell01}\end{center}
\end{figure}

Next we solve the BAE for $N=38$. This corresponds to the choice
$\delta_{b}=1$. The plots in Fig \ref{N19ell01} show that the slope
near the origin is again positive (modulo jumps) for $\gamma=\pi/2.5$
for smaller values of $\ell$.

\begin{figure}

\caption{The Bethe ansatz roots (left) and the corresponding $\Im m(\ln b)$
for $N=38,\gamma=\pi/3.5,\ell=0.1$}

\begin{center}\includegraphics{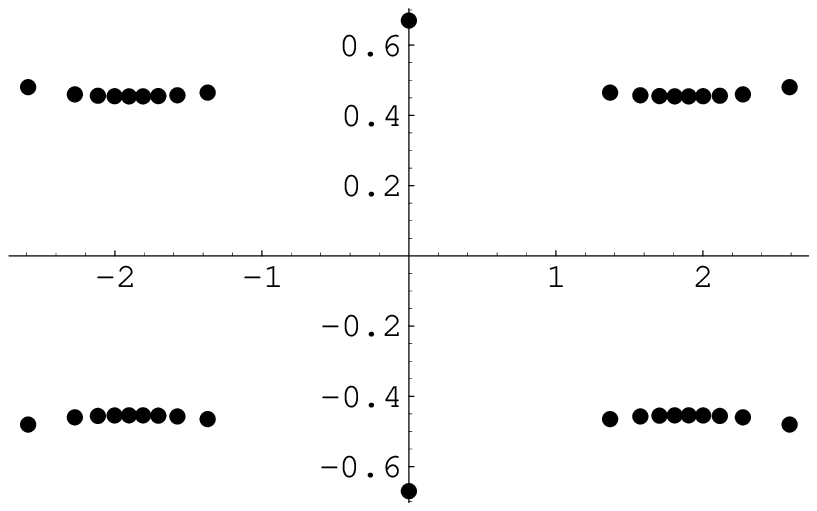}~~~~~\includegraphics{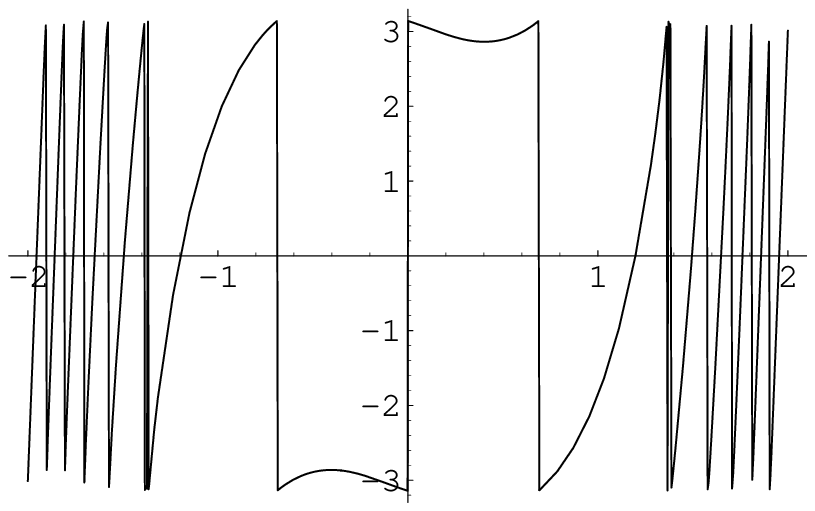}\label{N19ell01repulsive}\end{center}
\end{figure}

The plots in the repulsive regime ($\gamma=\pi/3.5$) are given in
Fig. \ref{N19ell01repulsive}. The right figure shows the negative
slope at the origin. (Imagine the {}``extended Brilliouin zone\char`\"{}
and line defined by $\Im mx=\pi$.) This indicates that the upper
component $\theta_{j}$ of the quasi 2-string centered at the origin
has $\left.\frac{d}{d\theta}\mathrm{Im}\ln b(\theta)\right|_{\theta=\theta_{j}}<0$,
i.e. it is a special object. While in the analog attractive case $N_{S}=0$
and for the vacuum where there are no holes $N_{eff}=N_{H}-2N_{s}=0$,
here we have now $N_{s}=1$. However, the jumps at $x\sim\pm0.7$
do not correspond to any pair of roots. Therefore they identify two
pairs of holes (they are \emph{not} special, as the derivative of
$\ln b$ is positive for them). In total we have now $N_{H}=2$, but
$N_{eff}$ continues to be 0.

\end{document}